\documentclass[11pt,twocolumn]{article}

\pdfoutput=1

\usepackage{graphicx}
\usepackage[hyphens]{url}

\begin{document}
\title{Tool Support of Formal Methods for Privacy by Design}
\author{Sibylle Schupp}
\date{}
\maketitle
\begin{abstract}
Formal methods are, in principle, suited for supporting  the recent
paradigm of privacy by design, but no overview is available 
that summarizes which particular approaches have been investigated, 
for which application domains they are suited, and whether they are 
implemented and available as tools. 
Using the techniques of search-based literature review and snowballing
this paper answers those questions for a selected set of research papers. 
\end{abstract}

\section{Introduction}
Whenever privacy regulations for software applications are formulated, they aim 
at making privacy an integral part of software development---a well-known 
slogan is ``privacy by design'', coined by 
Ann Cavoukian~\cite{cavoukian2011privacy}.
When private data is considered even critical, application developers might 
have to provide provable guarantees for respecting or implementing particular 
privacy concerns.  
Formal methods are, in principle, suited for supporting privacy by design, 
but no overview is available that summarizes which particular methods are 
``best practice'', or in use at all. 
The literature itself is spread and can be found not only in venues for 
applied formal methods, but also in conferences on software engineering, 
security, data management, or theory.

In this paper we provide a review of the current state of formal methods 
that may aid software development for privacy concerns. For each paper in a
selected list of papers we answer three questions:
\begin{enumerate}
\item[Q1.] Which particular formal method is presented?
\item[Q2.] In which application areas is this method applied and evaluated? 
\item[Q3.] Which tool---existing or newly developed---implements the            presented approach? 
\end{enumerate}

While the research area is certainly in flux, it is important to know for 
application developers as well as for judges or legal advisers whether obvious 
candidates of formal methods exist that either suggest some kind of “state of 
the art” or, conversely, render the use of particular other methods as 
negligent. 
This review can also be helpful for prospective researchers in the field, as 
there are only few conferences and journals dedicated to privacy, and 
formal-method papers are not necessarily published there. 

For selecting a set of primary studies, we employ the so-called hybrid 
approach, a method for systematic literature reviews 
that originates in software engineering. The hybrid approach consists of two stages, where 
stage 1 conducts a systematic search to obtain an initial list (\cite{kitchenham2004procedures,kitchenham2009systematic}) and stage 2, called snowballing (\cite{wohlin2014guidelines}),
traces the forward and backward references to and from that initial list. 
We explain the paper selection process in Section~\ref{sec:selection-tagging}; 
in this section we also detail how we pre-process the primary studies by
annotating them with sets of tags. 
The results themselves are presented in Section~\ref{sec:results}. 
Section~\ref{sec:threats} discusses threats to their
validity and Section~\ref{sec:con} concludes. 

\section{Selection and Tagging}\label{sec:selection-tagging}
In this section, we describe how the set of primary studies is 
constructed and how we pre-process those studies  by tagging them.

\subsection{Selection}\label{sec:selection}
\begin{figure}
\centerline{\includegraphics[scale=0.1]{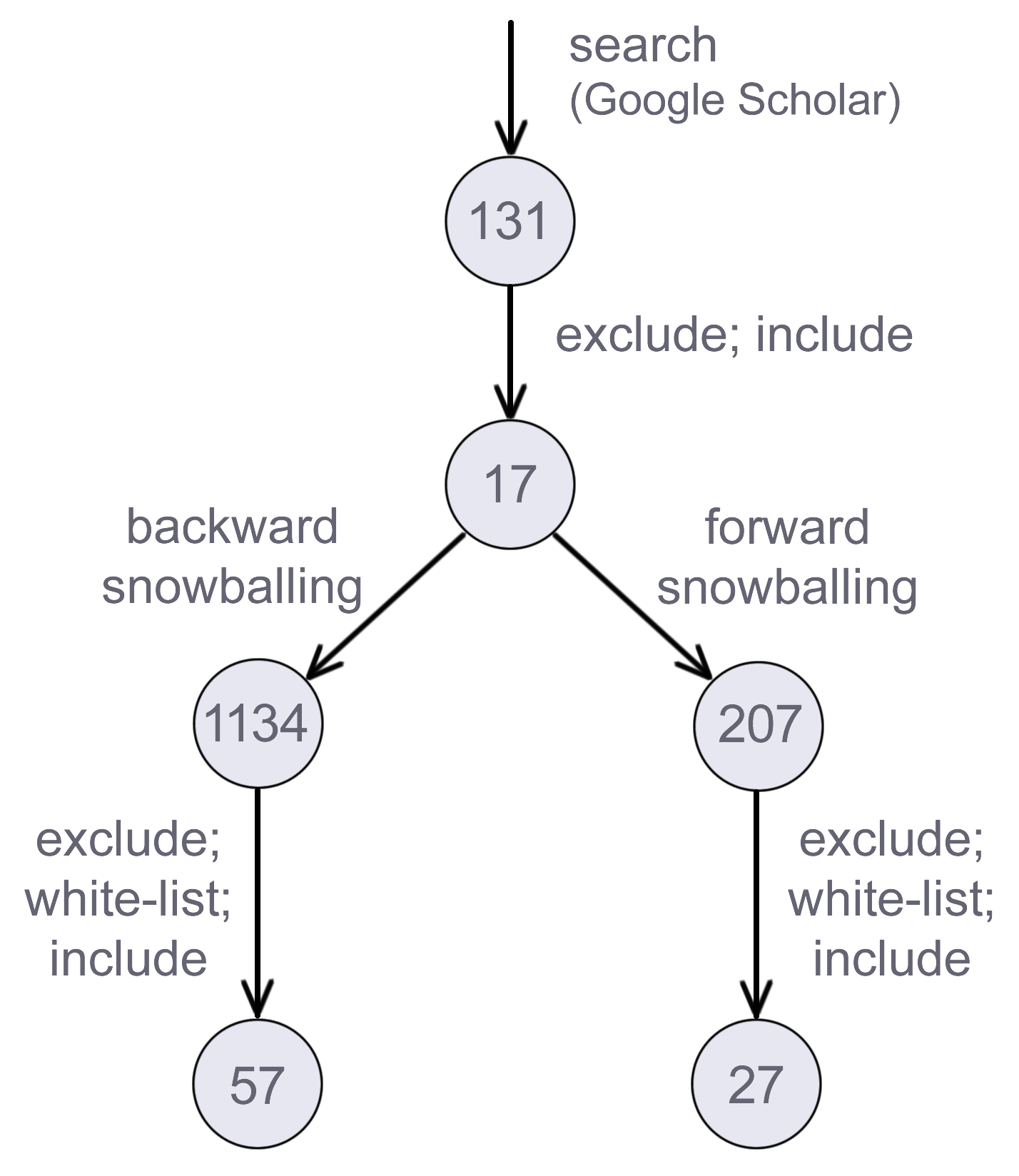}
}
\caption{Selection process: stage 1 (search), stage 2 (forward and backward snowballing), and resulting numbers of primary studies (17,57,27).}
\label{fig:Proc}   
\end{figure}

Following the hybrid approach for paper selection, we perform first a
search, then a snowballing step; for an overview, see 
Figure~\ref{fig:Proc}. 

In the search stage, we feed the search term (see Table~\ref{tab:SearchTerm}) 
to a search engine, Google scholar, and apply to the resulting hit list
a series of exclusion and inclusion criteria to filter out distractors. 
Those criteria correspond to common criteria for literature reviews (see
Table~\ref{tab:incex} in the appendix). Applying the criteria
 reduces the list from originally 131 papers to 17 papers. 
Those papers form the initial list, which serves as input to stage 2.

\begin{table}
\centering
\caption{Search term.} \vspace{1ex}
\label{tab:SearchTerm}     

\begin{tabular}{p{0.93\columnwidth}} 
\hline\noalign{\smallskip}
``privacy by design'' ``formal methods'' ``software engineering''  \\

\noalign{\smallskip}\hline

\centerline{\includegraphics[scale=0.35]{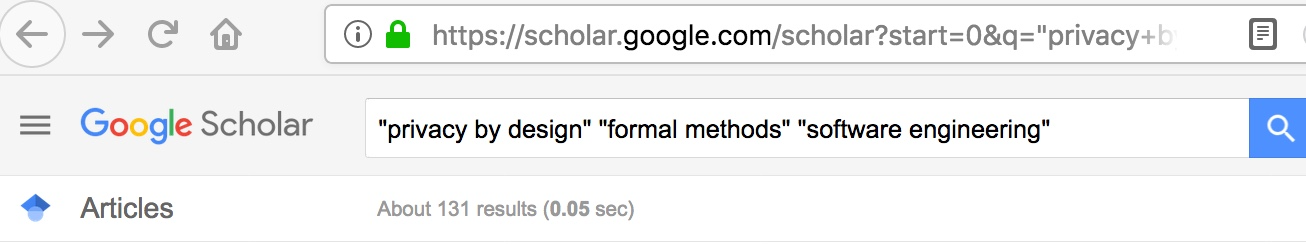}}
\end{tabular}
\end{table}

Stage 2 commences with backward snowballing, using the exclusion, inclusion, 
and white-listing criteria from before (see Table~\ref{tab:incex}). 
For each paper in the initial set, we go through its references and apply 
all exclusion criteria; on the remaining references we apply the white-list 
criteria to obtain a temporary list. 
Each reference on the white list, next, is then manually checked for each 
inclusion criterion  and, in the positive case, added to our list of primary 
studies. 
Forward snowballing proceeds similarly.  Using Google Scholar again,
we determine for each paper in the initial set all papers it is cited by. 
For each citing paper we repeat the process employed for backward snowballing, 
i.e., apply the exclusion criteria and the white-list criteria, and check 
by hand the remaining papers for topicality and availability according 
to the inclusion criteria. 

Table~\ref{tab:snow} provides the results. Altogether 57 additional papers can 
be obtained through backward snowballing and another 27 papers by forward 
snowballing.
For each paper in the initial set we list  the number of its
references (``refs'') for backward snowballing and citations (``cited'') for 
forward snowballing,  the number of potentially 
relevant papers (``inc?'') after white-listing, and the number of actually 
relevant (``inc!'') papers after the inclusion check.  
The last number refers to the papers that extend the initial set.

We note that the sum over those papers (row ``Sum'') is significantly larger
than the number of actually added papers (row ``Addit. papers''). This is no 
mistake but implies that some papers are cited multiple times. 
Multiple citations are a positive sign for the purpose of a review as they 
signify that a paper has a certain weight in the research discussion. 
We also note that the citation numbers (column ``cited'') are comparatively low.
The reason is that  the papers in the initial set are all very recent: 
8 (of the 17) papers have been published 2016 or later, and the remaining
ones date back to not later than 2011.

Altogether, this review is based on a set of 101 primary studies. 

\begin{table}
\centering
\caption{Selection process, stage 2 (snowballing): the initial set, relevant references (``backward''), relevant citations (``forward''); papers considered (``inc?'') and included (``inc!'').} 
\smallskip
\label{tab:snow}  
\begin{tabular}{p{0.12\columnwidth}|rrr|rrr}
\hline\noalign{\smallskip}
Initial            & \multicolumn{3}{l}{Backward} &  \multicolumn{3}{l}{Forward} \\
  set    & refs & inc? & inc! & cited & inc? & inc!\\
\noalign{\smallskip}\hline\noalign{\smallskip}
\cite{abe_formal_2016}   &   30    &   12      &    10    &   4     &   4       & 4 \\
\cite{akeel_secure_2017}    &  186 &    30     &   8     &  0       &   -       &  -\\

\cite{amato_model_2015}   & 46      & 2         & 1        & 37         & 10        & 0 \\
\cite{antignac_privacy_2014-1} & 34      & 18         & 6        & 18    & 10         & 10 \\
\cite{antignac_trust_2015}           & 34      & 21         & 6        & 8         & 4         & 2 \\
\cite{antignac_data_2016}         &   31    &  5        &  1       &  5       &   2      &  1 \\
\cite{benghabrit_abstract_2018}        &  66     &   16       &  12       &  0       &   -      &- \\
\cite{benthall_context_2018}        &   146    &   29       &   12      &  0       &   -      & - \\
\cite{chowdhury_policy_2018}          & 23      & 1         & 1        & 4         & 1         & 0 \\
\cite{kost_privacy_2011}          & 22      & 8        & 3        & 25        &  4        & 2 \\
\cite{kost_privacy_2012}          & 28      & 8        & 1        & 19        & 4         & 3 \\
\cite{le_metayer_privacy_2013}            & 53     & 20         & 10        & 22         & 14       & 13 \\
\cite{le_metayer_capacity:_2018}            &   17    &   7       &   6      &   0       & 0        & 0 \\
\cite{maffei_security_2013}            & 83      & 1         & 1        & 31        & 9        & 8 \\
\cite{pardo_formal_2014}          & 10     & 5        & 4       & 28      & 15         & 10 \\
\cite{shiyam_formal_2017}    &   303   &     6    &   0     &    0     &   -       &  -\\
\cite{ta_privacy_2015}    &   22    &   9      &   4    &  6       &  5        &  5\\
\hline\noalign{\smallskip}
Sum         & 1134 & 198 & 86  & 207 & 82 & 58 \\ \hline
Addit.  & & &   & & & \\
papers      & &  & 57  & &  & 27 \\ 
\hline \hline
%   & & & & & & 57 \\
%  & & & & & & 27 \\
%  & & & & & & 17 \\
  & & &  & \multicolumn{3}{r}{(initial set)  17} \\
 & & &  & \multicolumn{3}{r}{(backward)  57} \\
 & & &  & \multicolumn{3}{r}{(forward)  27} \\
Total & & & & & & 101 \\
\noalign{\smallskip}\hline
\end{tabular}
\end{table}

\subsection{Tagging}\label{sec:tagging}
With each of the three questions Q1-Q3 we associate a number of tags that are
used to answer that question. In each case, we introduce three kinds of tags. 
The first kind of tags is formed by  all related terms that are highlighted by 
the authors themselves, because they use them in the title, abstract, or 
keyword list of their paper.
Apart from minor grammatical changes, those terms are taken verbatim from
the papers and directly serve as tags. As it turns out, however, very little 
of the information relevant for our review questions is directly exposed in the title, abstract, or keyword list. We therefore read cursory through each paper and extend the list of tags by the terms found thereby. Those terms also form tags directly. 
The third kind of tags are meta-tags, which we introduce to make 
the absence of information explicit, to further classify papers, and 
to group tags; those latter tags are prefixed with ``meta-''. 
Table~\ref{tab:tags} contains the complete set of tags, ordered by 
associated question; all meta-tags
are listed in the last row.

\begin{table*}
\caption{Data tags and meta-tags.}
\label{tab:tags}   
\begin{tabular}{p{0.98\textwidth}}
\hline\noalign{\smallskip}
\textbf{Formal methods  (101):} % \\ 
 analysis, AND/OR tree, API, automata, 
automata-based, Bayesian networks, BDD, boolean logic, 
causal analysis, CTL-FO, code generator, colored Petri Net, compiler, conceptual model, data-dependence graph, 
data-flow diagram, data-flow graph, decision diagram, 
decision system, deducability, deontic logic, description logic, dynamic logic, epistemic logic, first-order logic, first-order relational logic, first-order temporal logic, formal definition, 
formal description, formal framework, formal method, formal model, formal notation, formal representation, formal step, formal verification, formalization, games, Gentzen calculus, Hoare, inference, inference system, information flow types, knowledge-based logic, language, logic, logic-based, logic of privacy, logic of privacy and utility, LTL, mapping, Markov-DP, modal logic, model checking, model, model-based, model-driven, modeling, 
model-theoretic semantics, ontology, pi-calculus,
policy analysis, policy automaton, P-RBAD
privacy calculus, probabilistic automata, probabilistic logic, process algebra, program analysis, proof, proof technique, 
quantitative approach, reasoning, 
relationship-based access control, 
representation, 
role-based access model, run-time verification, 
satisfiability, semantics, spatial context, specification, specification-declarative, specification language, specifying, state-space analysis, static analysis, static verification, symbolic execution, 
temporal constraints, temporal context, temporal logic, temporal operators, 
theorem prover, timed traces, 
transformation, translation, translator, types, type systems, weakest precondition, verification
\\
\\
\textbf{Application domains (38):} % \\
  bank-information-system, biometrics, cloud app, course-eval, 
delivery service, e-commerce,  electric vehicle charging, electronic health-record, e-learning, electronic service, ETP (electronic toll payment), employee data, financial services, health-care, health-information-system, healthcare-appraisal-system, ITS (intelligent transportation system), loan application, location data, loyalty systems, mobile app, 
passenger-name-record, pay-as-you-go, public-transportation-ticketing, 
regeneration system, review system, slippery road alert, smart grid, smart meter, social data, social network, spyware, tax preparation, transportation, university, voting, website-app, web shop
\\
\\
\textbf{Tool support (37):} % \\
Alloy, Caprice, CAPRIV, CAPVerDE, COMPASS, Couenne, CUDD,
eHealth Framework, event-B, Grok, IDP, Isabelle, Java, JIF, KeY,
Legalese, Margrave, Maude, mCRL2, MetaMORPH(h)SY, MSVL, 
MyHealth@Vanderbilt, Netlogo, PINQ, PrivaCIAS, ProVerif, 
ProZ, PV, SAIL, SemPref, S4P, Spin/Promela, 
TSPASS, unnamed tool, Uppaal, 
Z, Z3, ZQL
 \\
\\
\textbf{Meta-tags (27):}  no-app, no-tool, example, 
            extended-example, experimental, 
            meta-biometrics, 
       meta-e-commerce, meta-formal-model, meta-formal-other, meta-formal-verification, meta-health, meta-infsys,
       meta-language, meta-logic meta-model-checking, meta-ontology, 
       meta-process-algebra,  meta-semantics, meta-smart-grid,
       meta-social-network, meta-spec, meta-static-analysis, meta-transform,
       meta-transportation, meta-trust, meta-types, meta-web
\\ 
\noalign{\smallskip}\hline
\end{tabular}
\end{table*}

\section{Results}\label{sec:results}
We are now ready to present the findings. For each question Q1-Q3 we first 
explain how we aggregate the extracted tags, and then report the results in 
a bar chart, which tallies related papers and provides a numerical overview, 
and a table, which lists all related references explicitly. 

\subsection{Q1: Formal methods}
Since the number of tags associated with question Q1 totals 101 
(Table~\ref{tab:tags}), we need to group those tags as otherwise the 
presentation would become too fragmented. 
Table~\ref{tab:group-tags-q1} lists our partitioning into 13 groups where 
the group ``Other'' contains those tags that do not fit in any other cluster.
With the partitioning we intend to capture the view of the formal methods 
community, but our choice cannot be  free from subjectivity. 
Nor is the annotation process itself, which associates a paper with particular 
tags in the first place. 
We therefore decided to make our tagging transparent by publishing the 
bibliographic source file. Possible additional validity problems 
we discuss later, 
in Section~\ref{sec:threats}.

\begin{table*}[t]
\caption{Q1.  Formal methods: Grouping of tags.}
\label{tab:group-tags-q1}  
\begin{tabular}{p{0.2\textwidth}p{0.75\textwidth}} 
\hline\noalign{\smallskip}
\textit{Id} & \textit{Tags (see Table~\ref{tab:tags})} \\
Formal Model & formal description, formal framework, formal model, formal notation, formal representation, formal steps, formalization, model, model-based, model-driven, modeling, representation \\
Formal Verification & formal method, formal verification, proof, reasoning, static verification, verification \\
Language & language, specification language \\
Logic &  AND/OR tree, boolean logic, CTL-FO, decision diagram, deducibility, deontic logic, dynamic logic, epistemic logic, first-order logic, first-order relational logic, first-order temporal logic, Gentzen calculus, Hoare, inference, inference system, knowledge-based logic, LTL, logic, logic-based, logic of privacy,  logic of privacy and utility, modal logic, 
privacy calculus, probabilistic logic, proof technique, satisfiability, spatial context, symbolic execution, temporal logic, temporal operators, temporal constraints, temporal context, weakest precondition  
\\
Model Checking & model-checking, state-space analysis \\
Ontology & description logic, ontology \\
Process Algebra & process algebra, pi-calculus, colored Petri Net \\ 
Semantics & semantics, model-theoretic semantics \\
Specification &  declarative specification, formal definition, specification, 
                 specifying\\

Static Analysis & analysis, BDD, data-dependence graph, data-flow diagram, 
       data-flow graph, dependence graph, policy analysis, program analysis,
    static analysis  \\ 
Transform & code generator, compiler, mapping, transformation, translator,
translation \\ 
Types & information-flow types, types, type system \\
Other & API, automata, automata-based, causal analysis,
  conceptual-model, Bayesian networks, 
decision system, relationship-based access control,
         role-based access models, P-RBAC, theorem-prover, Markov-DP,  policy automaton, probabilistic automata, quantitative approach, 
       run-time verification, timed-traces
\\
\noalign{\smallskip}\hline\noalign{\smallskip}\\
\end{tabular}
\end{table*}

\begin{figure}
\includegraphics[width=1.0\columnwidth]{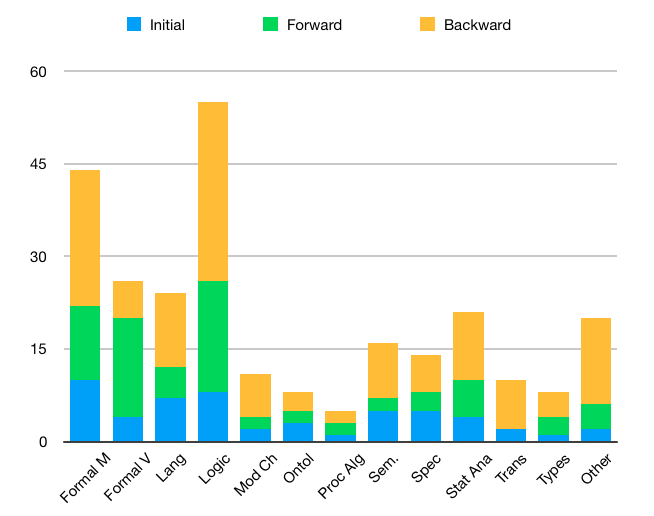}
\caption{Q1. Formal methods.}
\label{fig:fm-table} 

\end{figure}

Figure~\ref{fig:fm-table} reports the numbers for each group of formal methods,
sorted alphabetically, while Table~\ref{tab:q1-papers} lists for each 
group, or approach, of formal methods the papers that follow that approach. 
Interpreting the results from a practitioner's point of view, 
no obviously preferred formal method exists. 
Even the high numbers in the group ``Logic'' do not change the picture, as, 
upon  closer look,  the particular logics in this group vary greatly in 
their syntax, semantics, and pragmatics. 
Our second observation concerns the group ``Formal Model.'' Papers on formal 
methods often present their method directly since they can assume appropriate 
underlying formal representations (``models'').
For the domain of privacy, however, it is apparently  a research effort of 
its own to provide a formal model, and more than 40\% of the papers
is engaged in the some part of formalization.

\begin{table*}[t]
\caption{Q1.  Formal methods: listing by papers.}
\label{tab:q1-papers} 
\begin{tabular}{p{0.2\textwidth}p{0.75\textwidth}} 
\hline\noalign{\smallskip}
Formal Model & \cite{abe_formal_2016},
               \cite{agrafiotis_applying_2011},  \cite{akeel_secure_2017},
             \cite{amato_model_2015},
               \cite{antignac_privacy_2014-1}, \cite{antignac_trust_2015},
               \cite{barth_privacy_2006}, \cite{barth_privacy_2007},
               \cite{benthall_context_2018}, 
               \cite{bringer_privacy_2015}, \cite{bringer_reasoning_2015},
               \cite{bringer_reasoning_2016}, 
               \cite{datta_use_2017}, \cite{decroix_model-based_2015}, 
               \cite{decroix_framework_2013}, 
            \cite{deyoung_experiences_2010}, \cite{ding_model-driven_2010},
               \cite{fazouane_formal_2015}, \cite{fong_privacy_2009}, 
               \cite{fu_conformance_2010}, 
               \cite{hu_semantic_2011},
               \cite{kost_privacy_2012}, \cite{kost_privacy_2011},
               \cite{kouzapas_typing_2014},  
               \cite{krupa_handling_2012}, 
               \cite{le_metayer_formal_2008}, 
               \cite{le_metayer_capacity:_2018},
               \cite{li_privacy-aware_2009}, 
               \cite{liu_privacy-preserving_2017},
               \cite{liu_privacy-preserving_2016}, \cite{manna_towards_2018}, 
                \cite{moniruzzaman_delegation_2011},
              \cite{ni_conditional_2007}, \cite{ni_obligation_2008}, 
               \cite{ninghui_semantics-based_2006}, 
              \cite{omoronyia_engineering_2013},
              \cite{pardo_formalising_2017}, \cite{pardo_formal_2014}, 
              \cite{pardo_model_2017}, \cite{piolle_representing_2011}, 
              \cite{sun_privacy_2011}, \cite{suriadi_modeling_2009}, 
             \cite{tschantz_formalizing_2012}, 
              \cite{wang_method_2015}, \cite{zhang_modeling_2011}
 \\

Formal Verification & \cite{adams_constructing_2018},
                     \cite{agrafiotis_applying_2011}, 
                    \cite{amato_model_2015}, 
                      \cite{antignac_privacy-aware_2016},
                     \cite{antonakopoulou_ontology_2014},
                      \cite{backes_pricl:_2015}, 
                      \cite{benghabrit_checking_2015}, 
                     \cite{benghabrit_abstract_2018},
                     \cite{bringer_privacy_2015},
                     \cite{bringer_reasoning_2016}, 
                     \cite{bringer_reasoning_2015},
                     \cite{bringer_biometric_2017}, 
                  \cite{decroix_formal_2015}, \cite{decroix_framework_2013},                     \cite{fazouane_formal_2015}, 
                  \cite{fisler_verification_2005}, \cite{fu_conformance_2010}, 
                     \cite{kost_privacy_2011}, 
                   \cite{liu_privacy-preserving_2016},
                   \cite{omoronyia_caprice:_2012},  
                    \cite{pardo_formalising_2017}, 
                     \cite{pardo_timed_2018}, \cite{piolle_representing_2011},
                    \cite{ta_privacy_2018}, \cite{ta_privacy_2015},
                    \cite{wang_method_2015}  \\
Language & \cite{antignac_privacy_2014-1}, \cite{antignac_trust_2015},
           \cite{antignac_data_2016}, \cite{azraoui_-ppl:_2015},
           \cite{backes_automated_2012}, 
           \cite{bavendiek_privacy-preserving_2018},
          \cite{becker_s4p:_2010}, 
          \cite{becker_practical_2011}, \cite{benghabrit_accountability_2014},
          \cite{benghabrit_abstract_2018}, 
           \cite{chowdhury_policy_2018}, \cite{chowdhury_privacy_2013},
            \cite{fisler_verification_2005}, \cite{le_metayer_formal_2008}, 
          \cite{myers_protecting_2000}, \cite{ni_conditional_2007},
          \cite{ninghui_semantics-based_2006}, 
           \cite{pardo_formal_2014}, 
           \cite{pardo_specification_2016}, \cite{pardo_automata-based_2016},
          \cite{pitsiladis_implementation_2018},
          \cite{sen_bootstrapping_2014},
          \cite{ta_privacy_2018}, \cite{ta_privacy_2015}
\\
Logic &  \cite{agrafiotis_applying_2011}, 
         \cite{amato_model_2015}, \cite{anciaux_limiting_2012}, 
         \cite{antignac_privacy_2014-1},
          \cite{antignac_data_2016}, \cite{antonakopoulou_ontology_2014},
          \cite{aucher_dynamic_2011}, \cite{backes_unification_2004}, 
          \cite{backes_automated_2012}, \cite{barth_privacy_2006}, 
          \cite{backes_pricl:_2015}, 
         \cite{barth_privacy_2007},
        \cite{bavendiek_privacy-preserving_2018},
        \cite{becker_practical_2011}, \cite{benghabrit_accountability_2014},
            \cite{benghabrit_checking_2015}, \cite{benghabrit_abstract_2018},
         \cite{bringer_privacy_2015}, \cite{bringer_reasoning_2015},
         \cite{bringer_reasoning_2016},  
         \cite{bringer_biometric_2017}, \cite{chowdhury_privacy_2013},
         \cite{costante_privacy-aware_2013}, 
         \cite{datta_understanding_2011},  
         \cite{decroix_model-based_2015}, \cite{decroix_framework_2013},
         \cite{decroix_formal_2015}, 
        \cite{deyoung_experiences_2010},  \cite{fisler_verification_2005}, 
         \cite{fong_relationship-based_2011}, \cite{fu_conformance_2010}, 
         \cite{jafari_towards_2011},  \cite{karjoth_translating_2003}, 
         \cite{le_metayer_formal_2008}, 
         \cite{le_metayer_privacy_2013}, 
         \cite{kuang_security_2009},\cite{maffei_security_2013}, 
         \cite{manna_towards_2018},  \cite{moniruzzaman_delegation_2011},         \cite{ni_obligation_2008}, 
         \cite{ninghui_semantics-based_2006}, 
        \cite{omoronyia_caprice:_2012}, \cite{omoronyia_engineering_2013},
          \cite{pardo_formalising_2017}, 
          \cite{pardo_timed_2018}, \cite{pardo_formal_2014}, 
         \cite{pardo_model_2017},  \cite{pardo_specification_2016}, 
         \cite{pearson_decision_2011}, \cite{piolle_representing_2011}, 
         \cite{pitsiladis_implementation_2018},
         \cite{ta_privacy_2018}, \cite{ta_privacy_2015}, 
         \cite{tschantz_formal_2011},
        \cite{wang_method_2015} 
\\
Model Checking & \cite{abe_formal_2016}, \cite{amato_model_2015}, 
            \cite{aucher_dynamic_2011}, \cite{benghabrit_accountability_2014},
            \cite{fong_relationship-based_2011}, \cite{fu_conformance_2010},  
            \cite{jafari_towards_2011}, \cite{may_privacy_2006}, 
            \cite{pardo_model_2017},  \cite{pardo_specification_2016}, 
            \cite{suriadi_modeling_2009}  \\
Ontology & \cite{antonakopoulou_ontology_2014},
           \cite{barhamgi_privacy-preserving_2011},           
                     \cite{benthall_context_2018}, 
                     \cite{hecker_privacy_2008}, 
                 \cite{hu_semantic_2011},
                  \cite{knirsch_model-driven_2015}, 
                    \cite{kost_privacy_2012}, \cite{kost_privacy_2011}
\\
Process Algebra & \cite{fazouane_formal_2015}, \cite{kouzapas_typing_2014},
                  \cite{pitsiladis_implementation_2018},   
                  \cite{ta_privacy_2015} \\
Semantics & \cite{antignac_privacy_2014-1}, \cite{antignac_trust_2015},
            \cite{antignac_data_2016},  \cite{becker_s4p:_2010}, 
            \cite{becker_practical_2011}, 
            \cite{benghabrit_abstract_2015}, \cite{benghabrit_abstract_2018},
            \cite{datta_privacy_2014}, \cite{datta_understanding_2011}, 
            \cite{jafari_towards_2011}, 
            \cite{kost_privacy_2012}, \cite{ninghui_semantics-based_2006}, 
           \cite{pardo_formalising_2017},  \cite{sen_bootstrapping_2014},
           \cite{ta_privacy_2018}, \cite{tschantz_formalizing_2012} 
\\
Specification &  \cite{antignac_privacy_2014-1}, 
                 \cite{antignac_data_2016},
                 \cite{becker_s4p:_2010}, 
                 \cite{benghabrit_abstract_2018},
                 \cite{bringer_biometric_2017},
                 \cite{chowdhury_privacy_2013}, 
                  \cite{datta_understanding_2011}, 
                 \cite{deyoung_experiences_2010}, 
                  \cite{fu_conformance_2010}, 
                 \cite{maffei_security_2013},
                 \cite{pardo_automata-based_2016},
                  \cite{sen_bootstrapping_2014},
                 \cite{shiyam_formal_2017}, \cite{ta_privacy_2018} 
\\
Static Analysis & \cite{antignac_privacy-aware_2016},
                  \cite{chowdhury_privacy_2013},
                  \cite{datta_use_2017},                
                  \cite{decroix_model-based_2015},
                  \cite{decroix_formal_2015},  \cite{fisler_verification_2005}, 
                   \cite{fong_privacy_2009},
                  \cite{fu_conformance_2010}, 
                  \cite{knirsch_model-driven_2015}, \cite{knirsch_privacy_2015},
                  \cite{kost_privacy_2012}, \cite{kost_privacy_2011},
                  \cite{liu_privacy-preserving_2016},
                  \cite{maffei_security_2013}, \cite{myers_protecting_2000}, 
                   \cite{sen_bootstrapping_2014},
                  \cite{shiyam_formal_2017},  \cite{suriadi_modeling_2009},
                  \cite{tierney_realizing_2014}, 
                   \cite{zanioli_sails:_2012}, \cite{zhang_modeling_2011}
\\
Transformation & \cite{amato_model_2015}, \cite{azraoui_-ppl:_2015},
                  \cite{backes_unification_2004}, 
                  \cite{backes_automated_2012}, 
                 \cite{benghabrit_abstract_2015}, 
                 \cite{ding_model-driven_2010}, 
                 \cite{fournet_zql:_2013}, \cite{karjoth_translating_2003}, 
                 \cite{mcsherry_privacy_2009},
          \cite{ta_privacy_2015} \\
Types & \cite{adams_constructing_2018},
        \cite{cortier_type_2017}, \cite{decroix_framework_2013}, 
        \cite{kouzapas_typing_2014},  \cite{maffei_security_2013},
        \cite{myers_protecting_2000}, 
        \cite{pitsiladis_implementation_2018}, 
        \cite{sen_bootstrapping_2014}\\
Other & 
          \cite{antignac_privacy-aware_2016} (conceptual model), 
          \cite{benghabrit_checking_2015} (theorem prover),   
          \cite{benthall_context_2018} (Bayesian networks),
          \cite{datta_privacy_2014} (Markov-DP), 
           \cite{datta_use_2017} (causal analysis),
           \cite{fong_relationship-based_2011} (relation-based),           
        \cite{kuang_security_2009} (role-based access model),
\cite{maffei_security_2013} (API), 
          \cite{moniruzzaman_delegation_2011} (P-RBAC),
          \cite{ni_conditional_2007} (role-based),
          \cite{ni_obligation_2008} (role-based), 
         \cite{pace_runtime_2016} (automata-based, run-time verification), 
          \cite{pardo_timed_2018} (timed-traces),
          \cite{pardo_automata-based_2016} (policy automaton,
        run-time verification),
          \cite{pearson_decision_2011} (decision system),  
       \cite{pinisetty_monitoring_2018} (run-time verification),      
      \cite{tschantz_formal_2011} (prob. automata),
      \cite{tschantz_formalizing_2012} (Markov-DP),
     \cite{tschantz_purpose_2013} (Markov-DP),      
       \cite{zhang_modeling_2011} (conceptual model, quant. approach)\\
      
\noalign{\smallskip}\hline
\end{tabular}
\end{table*}

\subsection{Q2. Application domains}
For presenting the targeted application domains we, again, we partition the 
corresponding tags. In this case, we partition 38 tags 
(see Table~\ref{tab:tags}) 
in 9 groups, using the corresponding application domain as grouping criterion 
(see Table~\ref{tab:group-tags-q2}). 
Most group memberships are easy to decide syntactically and the application 
domains are largely as expected (see, e.g., the groups ``E-Commerce'', 
``Health'', ``Smart Grid''). 

\begin{table*}[h]
\caption{Q2. Application domains: Grouping of tags.}
\label{tab:group-tags-q2}   
\begin{tabular}{lp{0.74\textwidth}}
\hline\noalign{\smallskip}
\textit{Id}        & \textit{Tags (see Table~\ref{tab:tags}) }\\
Biometrics & biometrics \\
E-Commerce &e-commerce, electronic service, delivery service,              
            loyalty systems, passenger-name-record, 
             public-transportation-ticketing, web shop\\
Health & electronic health-record, healthcare-appraisal-system, health-care,
              health-information-system \\
Information Systems & bank-information-system, 
                      course-eval, e-learning, employee data, 
                      financial services, 
                     loan application, review system, university \\ 
Transportation & electric vehicle charging, ETP (electronic toll payment), 
                ITS (intelligent transportation system), location service, pay-as-you-go, transportation\\
Smart Grid & smart grid, smart meter\\
Social Network & social data, social network \\
Web   & cloud-app, mobile ads, mobile app, website-app \\
Other         & GUI-design, regeneration system, slippery-road alert, spyware, tax preparation,
                voting \\
\hline\noalign{\smallskip}
\end{tabular}
\end{table*}

The majority of the papers include examples mostly for the purpose of 
illustrating a theoretical point. Two important tags are therefore the 
meta-tags ``example-extended'' and ``experimental'', which flag extended, 
(and ideally) realistic examples and empirical measurements, thus mark 
those papers that are of particular interest to application developers.

\begin{figure}

 \includegraphics[width=1.0\columnwidth]{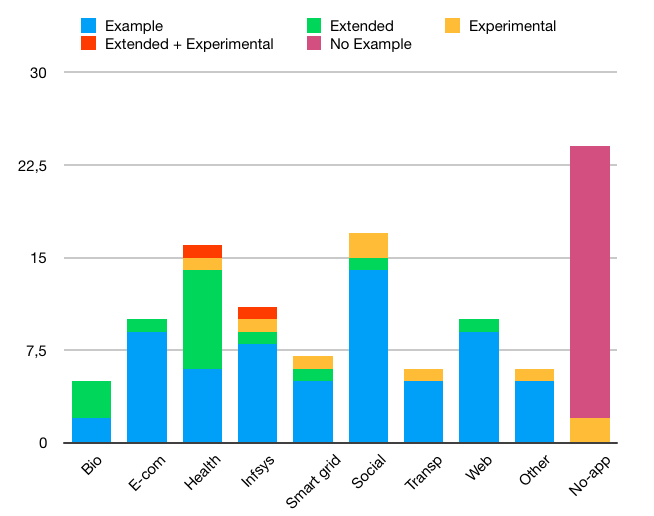}
\caption{Q2. Application domains.}
\label{fig:q2-table}     
\end{figure}

Again, we answer the question by providing a bar chart and a table: 
Figure~\ref{fig:q2-table}  shows the distribution of papers
across application domains as well as the distribution of papers containing
extended examples or experimental data. Conversely, Table~\ref{tab:q2-papers} 
lists for each application domain the papers that relate to it. %

Comparing the numbers in Figure~\ref{fig:q2-table}, one can see 
two application domains with higher numbers: ``Health'' and ``Social Networks.'' 
Looking closer at the numbers of  ``extended examples'' or 
``extended + empirical'', the domain ``Health'' clearly exhibits the largest
number of papers of practical relevance. 
Overall, however, the ratio of papers with extended examples is very low, 
amounting to less than a fifth. 
At the same time, the dominating number in the figure is the number 
of papers without examples beyond illustrative snippets (``No-app''), 
just a bit more than a fifth.
Altogether, it is fair to say that---with the possible exception of the health domain---current formal approaches cannot yet refer to many showcases. 
Accordingly, experimental evaluations---as they matter for practical
software development---are scarce: only 9 papers
contain experimental data and only two papers contain both an extended example
and its empirical evaluation.  The bar ``No-app'' includes additional two 
papers with experimental data: those papers measure
performance parameters of the presented tool rather than  an application.

For a legal advisor those numbers imply that there are no ``best'' formal
methods  one can application developers expect to follow; the only domain 
where it is sensible at all to ask for best practices is the health-care 
domain.  For application developers those numbers mean that, at present, there
is little software to reuse. 

\begin{table*}[h]
\caption{Q2. Application domains: listing by papers. ``X'' marks papers with extended examples, ``E'' papers with experimental data.}
\label{tab:q2-papers}
\begin{tabular}{lp{0.74\textwidth}}
\hline\noalign{\smallskip}
Biometrics & \cite{benthall_context_2018}, \cite{bringer_privacy_2015} X,
             \cite{bringer_reasoning_2016} X, \cite{bringer_biometric_2017} X
\\
E-Commerce &  
              \cite{antignac_data_2016}, 
              \cite{benthall_context_2018}, 
              \cite{costante_privacy-aware_2013} X, 
              \cite{decroix_model-based_2015},
              \cite{decroix_framework_2013},
              \cite{decroix_formal_2015}, 
                          \cite{hecker_privacy_2008}, 
              \cite{hu_semantic_2011},
               \cite{karjoth_translating_2003}, \cite{pinisetty_monitoring_2018}
\\
Health & \cite{akeel_secure_2017} X, 
              \cite{amato_model_2015} X, 
              \cite{barhamgi_privacy-preserving_2011} EX,           
              \cite{barth_privacy_2007} X, 
              \cite{benghabrit_abstract_2015} X, 
              \cite{benghabrit_checking_2015} X, 
              \cite{chowdhury_privacy_2013} X,
              \cite{ding_model-driven_2010},
              \cite{fong_relationship-based_2011} X,
              \cite{hu_semantic_2011},    
              \cite{jafari_towards_2011}, 
              \cite{kouzapas_typing_2014},               
              \cite{kuang_security_2009}, 
              \cite{shiyam_formal_2017} X,
              \cite{tschantz_purpose_2013}, 
              \cite{tschantz_formalizing_2012} E
\\
Information Systems & \cite{agrafiotis_applying_2011},
                      \cite{anciaux_limiting_2012} E, 
                      \cite{backes_automated_2012}, 
                      \cite{chowdhury_policy_2018},
                       \cite{datta_use_2017} E,                 
                      \cite{decroix_framework_2013},
                      \cite{fisler_verification_2005}, 
                     \cite{fong_privacy_2009}, 
                     \cite{maffei_security_2013} EX
                      \cite{manna_towards_2018}, 
                          \cite{piolle_representing_2011}, 
                      \cite{shiyam_formal_2017} X
\\
Smart Grid & \cite{abe_formal_2016},
              \cite{antignac_privacy_2014-1}, 
             \cite{bavendiek_privacy-preserving_2018}, 
             \cite{fournet_zql:_2013} E,
             \cite{knirsch_model-driven_2015}, \cite{knirsch_privacy_2015}, 
             \cite{ta_privacy_2018} X
\\
Social Networks & \cite{backes_automated_2012}, 
                 \cite{fong_relationship-based_2011} X, 
                 \cite{fong_privacy_2009},
                 \cite{kouzapas_typing_2014},  
                   \cite{le_metayer_capacity:_2018},
              \cite{omoronyia_engineering_2013} E, \cite{pace_runtime_2016}, 
                 \cite{pardo_formalising_2017}, 
                \cite{pardo_timed_2018}, \cite{pardo_formal_2014},
                \cite{pardo_model_2017},  \cite{pardo_specification_2016}, 
                \cite{pardo_automata-based_2016}, \cite{wang_method_2015}\\
Transportation & \cite{antignac_trust_2015}, 
     \cite{fazouane_formal_2015}, \cite{fournet_zql:_2013} E, 
    \cite{kost_privacy_2012},
       \cite{kost_privacy_2011},  \cite{krupa_handling_2012}, 
      \cite{le_metayer_privacy_2013},
       \cite{tierney_realizing_2014}
\\
Web  & \cite{adams_constructing_2018}, \cite{aucher_dynamic_2011}, 
               \cite{becker_s4p:_2010}, \cite{becker_practical_2011}, 
              \cite{benghabrit_accountability_2014},
             \cite{fu_conformance_2010}, 
             \cite{liu_privacy-preserving_2016} X,
             \cite{liu_privacy-preserving_2017},
             \cite{ni_obligation_2008}, \cite{omoronyia_caprice:_2012} \\
%\\
Other & \cite{antignac_privacy-aware_2016} (slippery-road alert),
        \cite{aucher_dynamic_2011} (spyware), 
        \cite{cortier_type_2017} E (voting), 
        \cite{myers_protecting_2000} (tax preparation), 
      \cite{pearson_decision_2011} (GUI design),
     \cite{zhang_modeling_2011} (regeneration system)
\\
no-app & \cite{antonakopoulou_ontology_2014}, \cite{azraoui_-ppl:_2015}, 
          \cite{backes_unification_2004},  \cite{backes_pricl:_2015}, 
         \cite{barth_privacy_2006}, \cite{benghabrit_abstract_2018}, 
         \cite{datta_privacy_2014},
         \cite{datta_understanding_2011}, \cite{deyoung_experiences_2010}, 
         \cite{karjoth_translating_2003}, 
         \cite{le_metayer_formal_2008}, \cite{li_privacy-aware_2009}, 
         \cite{may_privacy_2006}, \cite{mcsherry_privacy_2009},
         \cite{moniruzzaman_delegation_2011}, \cite{ni_conditional_2007},
         \cite{ninghui_semantics-based_2006}, \cite{pitsiladis_implementation_2018},  \cite{sen_bootstrapping_2014} E,
        \cite{sun_privacy_2011}, \cite{suriadi_modeling_2009}, 
         \cite{ta_privacy_2015}, \cite{tschantz_formal_2011}, 
         \cite{zanioli_sails:_2012} E\\

\hline\noalign{\smallskip}
\end{tabular}
\end{table*}

\subsection{Q3. Tool support}
\begin{figure}
  \includegraphics[width=1.0\columnwidth]{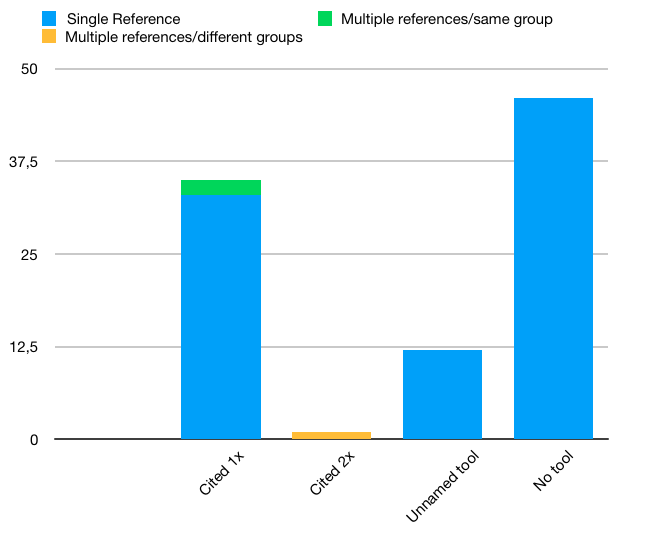}
\caption{Q3. Tool support.}
\label{fig:q3-table}   
\end{figure}

The last question concerns the tool a paper either uses to implement its 
approach or newly develops for that purpose. 
Figure~\ref{fig:q3-table} summarizes the current tool support
by breaking the set of tools down in three categories: unnamed tools, coming
from papers that devise a prototypical implementation; (named) tools that are
used either in one paper only or, if cited multiple times, by the same research
group; and, lastly, tools that are used by different research groups. 
As forth category, we include those papers that mention no tool. 
In Table~\ref{tab:q3-papers} we list which tool implements which approach.

The results are perhaps unexpected. For one, one might expect the
overall number of tools to be higher---after all the primary studies 
are authored by software engineers or computer scientists. Yet, almost half 
of the studies is purely theoretical in that sense (``No tool'' in 
Figure~\ref{fig:q3-table}). 
The spreading of tools, second, is very large: only four tools are used in 
more than one paper (see Table~\ref{tab:q3-papers}). 
Generally, such spreading of tools is quite common for formal methods and can 
be seen elsewhere as well. 
For the case at hand, high numbers follow already from the variety of formal 
methods used and the variety within those groups: the group ``Logic'', for 
example comprises a wide range of logics, for which no single tool exists. 
Still, it is remarkable that there is only one tool in our primary studies,
Alloy, that is used by  more than one research group.

\begin{table*}[h]
\caption{Q3. Tool support: listing by papers.}
\label{tab:q3-papers}
\begin{tabular}{lp{0.74\textwidth}}
\hline\noalign{\smallskip}
\multicolumn{2}{l}{
\begin{tabular}{p{0.15\textwidth}p{0.23\textwidth}p{0.25\textwidth}p{0.23\textwidth}}
Alloy & \cite{chowdhury_policy_2018}, \cite{fu_conformance_2010} & 
Caprice & \cite{omoronyia_caprice:_2012} \\
 CAPRIV & \cite{antignac_trust_2015}& 
CAPVerDE & \cite{bavendiek_privacy-preserving_2018} \\
COMPASS  & \cite{tierney_realizing_2014} &
Couenne  & \cite{anciaux_limiting_2012} \\
CUDD     &  \cite{fisler_verification_2005} &
eHealth Framework & \cite{ding_model-driven_2010} \\
event-B & \cite{akeel_secure_2017} &
Grok  &  \cite{sen_bootstrapping_2014} \\
IDP   & \cite{decroix_model-based_2015} (knowledge-based system)&
Isabelle & \cite{shiyam_formal_2017} \\
Java     & \cite{costante_privacy-aware_2013}&
JIF      & \cite{myers_protecting_2000} \\
KeY & \cite{antignac_data_2016} &
Legalese &  \cite{sen_bootstrapping_2014} \\
Margrave &  \cite{fisler_verification_2005} &
Maude & \cite{pitsiladis_implementation_2018}, \\
mCRL2 & \cite{benghabrit_accountability_2014}&
MetaMORPH(h)SY & \cite{amato_model_2015}\\
MSVL & \cite{wang_method_2015} &
MyHealth@Vanderbilt & \cite{barth_privacy_2007} \\
Netlogo & \cite{omoronyia_engineering_2013} &
PINQ  & \cite{mcsherry_privacy_2009} \\
PrivaCIAS & \cite{krupa_handling_2012} &
Proverif & \cite{fazouane_formal_2015} \\
ProZ  & \cite{liu_privacy-preserving_2016} &
PV & \cite{fu_conformance_2010} \\
SAIL & \cite{zanioli_sails:_2012} & 
SemPref & \cite{ninghui_semantics-based_2006} (language) \\
S4P &  \cite{becker_s4p:_2010}, \cite{becker_practical_2011} (language) & 
 Spin/Promela & \cite{may_privacy_2006} \\
TSPASS & \cite{benghabrit_abstract_2015}, \cite{benghabrit_checking_2015}, 
         \cite{benghabrit_abstract_2018}        &
 unnamed tool &  \cite{antignac_data_2016}, \cite{backes_unification_2004}, 
         \cite{backes_automated_2012}, \cite{benghabrit_abstract_2018},                 \cite{cortier_type_2017},  \cite{datta_use_2017},                
            \cite{knirsch_privacy_2015}, 
            \cite{maffei_security_2013}, 
           \cite{pearson_decision_2011},
           \cite{sen_bootstrapping_2014}, \cite{tschantz_formalizing_2012},
          \cite{zhang_modeling_2011} \\ 
Uppaal &  \cite{amato_model_2015}& 
Z & \cite{abe_formal_2016}, \cite{liu_privacy-preserving_2016} \\
Z3 & \cite{antignac_data_2016} & 
ZQL & \cite{fournet_zql:_2013} \\
\\
no-tool & \cite{adams_constructing_2018}, \cite{agrafiotis_applying_2011},
          \cite{antignac_privacy-aware_2016}, 
         \cite{antonakopoulou_ontology_2014}, \cite{aucher_dynamic_2011}, 
         \cite{azraoui_-ppl:_2015}, \cite{backes_pricl:_2015}, 
         \cite{barhamgi_privacy-preserving_2011},           
         \cite{barth_privacy_2006}, 
         \cite{bringer_privacy_2015}, \cite{bringer_reasoning_2015},
         \cite{bringer_reasoning_2016}, 
         \cite{bringer_biometric_2017}, \cite{chowdhury_privacy_2013},
         \cite{datta_privacy_2014}, \cite{datta_understanding_2011}, 
        \cite{decroix_formal_2015}, 
     &  \cite{decroix_framework_2013}, 
        \cite{deyoung_experiences_2010}, 
        \cite{fong_privacy_2009},
        \cite{fong_relationship-based_2011}, 
        \cite{hecker_privacy_2008},  \cite{hu_semantic_2011},
       \cite{jafari_towards_2011}, 
        \cite{karjoth_translating_2003}, 
        \cite{knirsch_model-driven_2015}, \cite{kouzapas_typing_2014}, 
        \cite{kuang_security_2009},  
        \cite{le_metayer_formal_2008}, \cite{li_privacy-aware_2009}, 
       \cite{moniruzzaman_delegation_2011}, 
        \cite{ni_conditional_2007}, \cite{ni_obligation_2008},         
        \cite{pace_runtime_2016}, \cite{pardo_formalising_2017}, 
        \cite{pardo_timed_2018}, 
       \cite{pardo_formal_2014}, 
  & \cite{pardo_model_2017},  
        \cite{pardo_specification_2016}, \cite{pardo_automata-based_2016},
        \cite{pinisetty_monitoring_2018}, 
       \cite{piolle_representing_2011},           
          \cite{sun_privacy_2011}, 
        \cite{suriadi_modeling_2009},  \cite{ta_privacy_2018},
        \cite{tschantz_formal_2011}, \cite{tschantz_purpose_2013} 
 \\
\hline\noalign{\smallskip}

\end{tabular}
}

\end{tabular}
\end{table*}

\section{Threats to Validity}
\label{sec:threats}
The results of
this review depend critically on the tags each paper is characterized by,
but the process of tagging is done manually, and this presents a threat
to the validity of the results. In this subsection we first explain how we
address this threat, then discuss additional potential validity problems. 
 
Unfortunately, tagging by hand is inevitable---the majority of
relevant information is not directly exposed to the readers but ``hidden''
in the text. We double-checked so that we can ensure that the set of tags for terms that 
the authors themselves consider relevant (by including them in the title, 
abstract, or keyword list of their paper) is complete. 
Given the number of papers processed, however, we 
cannot dismiss the risk of having overlooked tags 
that emerge from cursory reading. Nor can we give a formal argument for
our specific partitioning of tags. 
The best we can do for mitigating validity problems is to make
our groupings transparent. We do so through the introduction
of appropriate meta-tags, and store with each paper its associated
meta-tags. 

The second source of threats to the overall validity concerns the selection 
process. A different search term as well as a different search engine would 
lead to a different initial list, while an additional 
iteration of the snowballing step would lead to
an extended list of primary studies. On the other hand, research for formal
methods on privacy has started only and we look at the references in this
review  as a starting point only, for a list that  needs to be extended as 
research moves on. We do not dispute that relevant papers might be missing
in the current review but we are confident that, using the technique of 
snowballing, one ultimately reaches all relevant papers.

While it will not be possible to exclude all mistakes and subjective 
assessments, we can at least ensure the full reproducibility of the results
and therefore make the complete data set available: the bibtex information
 for the primary
studies can be downloaded as three separate files\footnote{{\url{https://www.tuhh.de/sts/research/data-protection-machine -learning/initial17.bib} (\url{include-forward.bib}, \url{include-backward.bib})}}; all tags and meta-tags can be found
in the field ``keywords.'' 
The files were generated using the open-source software Zotero for reference 
management \cite{zotero}, which allows for various filters and conversions 
to spreadsheet and bibtex formats, and which was of great help in automating 
the evaluation.  

\section{Summary and Conclusion}
\label{sec:con}
Since personal data can be considered critical information,  
application developers may want to employ formals methods so that
they can argue in a rigorous manner that private data is protected properly. 
At present, however, no dedicated venues
exist where formal methods are discussed  that target specifically privacy 
concerns. Instead, the literature is scattered over conferences for
formal methods, software engineering, security, data management, or theoretical computer
science. In this paper we provide a systematic review of the use of
formal methods for privacy protection in practice.

 A wide variety of methods is available already today, and there
is no evidence for more, or less, suitable methods. Notably, a large 
portion of papers falls in the categories ``Formal Model'' and ``Language'': 
since privacy regulations are provided in natural language, extra modeling 
efforts are needed. 
On the other hand, comparatively few examples are available that application 
developers could take as a blueprint. Also, about half of the papers do not 
present an implementation of their approach and, generally, dedicated tool 
support is in its infancy. 

Concluding, privacy protection is an important topic where formal methods
can contribute in significant ways. As the publication years of the primary 
studies show, research in this field has gained momentum. 
As the review also shows, tooling support is a worthwhile subject for 
further research. 

\paragraph{Acknowledgment}
My sincere thanks are to Wolfgang Schulz, Max von Grafenstein,
 J{\"o}rg Pohle, and the Humboldt Institute for Internet 
and Society (HIIG) for feedback and discussions on the legal issues of privacy. 
I also owe to the participants of the Dagstuhl seminar 18471, "Next Generation Domain Specific Conceptual Modeling: Principles and Methods",
and to the Dagstuhl library for granting access to all literature
needed for this review. 

The paper is part of the project ``Information Governance Technologies: Ethics, Policies, Architectures, Engineering'' funded by the ``Landesforschungsf{\"o}rderung Hamburg.''
\appendix

\begin{table}
\caption{Criteria for inclusion, exclusion, and white-listing.\vspace{1ex}} 
\label{tab:incex}       

\begin{tabular}{lp{0.8\columnwidth}}
\hline\noalign{\smallskip}
\multicolumn{2}{l}{Inclusion criteria}\\
I1 & Papers meeting the search terms ``privacy'', ``formal method'', ``software'' and written for a  computer science/software engineering audience    \\
I2 & Papers available in full text\\
\noalign{\smallskip}\hline\noalign{\smallskip}%\\
\multicolumn{2}{l}{Exclusion criteria}\\
E1 & Handbooks, collections, surveys, position papers, milestone report, project summaries, master theses  \\
E2 & Standards, manuals, guidelines\\
E3 & Papers not written in English  \\
E4 & Duplicates, including short versions of included papers  \\
E5 & Papers published in law journals or conferences \\
E6 & Artifact links, e.g., to web-pages instead of papers \\
E7 & Papers dealing with formal methods in domains other than privacy \\
E8 & Papers dealing with mathematical/computer science methods for privacy concerns other than formal methods   \\
\noalign{\smallskip}\hline\noalign{\smallskip}%\\
\multicolumn{2}{l}{White-list criteria (snowballing)} \\ 
W1 & Papers with relevant title \\
W2 & Papers by authors already included \\
W3 & Papers in relevant conferences or journals\\
\noalign{\smallskip}\hline\noalign{\smallskip}\\
\end{tabular}
\end{table}

\sloppy
% \printbibliography  % biber command
\bibliography{toolSupportFormalMethodsPrivacy}{}
\bibliographystyle{plain}

\end{document}